# Dynamically Triangulated Ising Spins in Flat Space

*Marco Vekić and Shao Liu*

Department of Physics, UC Irvine
Irvine, Ca 92717, USA

*and*

*Herbert W. Hamber*

Theory Division, CERN
CH-1211 Genève 23, Switzerland

## ABSTRACT

A model describing Ising spins with short range interactions moving randomly in a plane is considered. In the presence of a hard core repulsion, which prevents the Ising spins from overlapping, the model is analogous to a dynamically triangulated Ising model with spins constrained to move on a flat surface. It is found that as a function of coupling strength and hard core repulsion the model exhibits multicritical behavior, with first and second order transition lines terminating at a tricritical point. The thermal and magnetic exponents computed at the tricritical point are consistent with the exact two-matrix model solution of the random Ising model, introduced previously to describe the effects of fluctuating geometries.

# 1  Introduction

Following the exact solution of the Ising model on a random surface by matrix model methods [1], there has been a growing interest in the properties of random Ising spins coupled to two-dimensional gravity. More recently, work based on both series expansions [2] and numerical simulations [3, 4] has verified and extended the original results. It is characteristic of these Ising models that the spins are allowed to move at random on a discretized version of a fluid surface. In a specific implementation of the model, Ising spins are placed at the vertices of a lattice built out of equilateral triangles, and the lattice geometry is then allowed to fluctuate by varying the local coordination number through a "link flip" operation which varies the local connectivity [3]. Remarkably the same critical exponents have also been found using consistency conditions derived from conformal field theory for central charge $c = \frac{1}{2}$ [5], which should again apply to Ising spins. It is generally believed that the new values for the Ising critical exponents are due to the random fluctuations of the surface (or the world sheet in string terminology) in which the spins are embedded, and therefore intimately tied to the intrinsic fractal properties of fluctuating geometries. It came therefore as a surprise that non-random Ising spins, placed on a randomly fluctuating geometry but with fixed spin coordination number, exhibited the same critical behavior as in flat space, without any observed "gravitational" shift of the exponents [6].

The natural question is then to what extent the values of the critical Ising exponents found in the matrix model solution ($\alpha = -1$, $\beta = 1/2$, $\gamma = 2$, $\eta = 2/3$, $\nu = 3/2$ [1]) are due to the *annealed* randomness of the lattice, and to what extent they are due to the physical presence of a fluctuating background metric. The most straightforward way to answer this question is to investigate the critical properties of annealed random Ising spins, with interactions designed to mimic as closely as possible the dynamical triangulation model, but placed in flat two-dimensional space. We should add that it is well known that for a *quenched* random lattice the critical exponents are the same as on a regular lattice [7], as expected on the basis of universality, even though in two dimensions the Harris criterion (which applies to quenched impurities only) does not give a clear prediction, since the specific heat exponent vanishes, $\alpha = 0$, for Onsager's solution.

In this letter we present some first results concerning the exponents of such a model. A more detailed account of our results will be the subject of a forthcoming publication [11].

## 2 Formulation of the Model

In a square $d$-dimensional box of sides $L$ with periodic boundary conditions we introduce a set of $N = L^d$ Ising spins $S_i = \pm 1$ with coordinates $x_i^a$, $i = 1...N$, $a = 1...d$, and average density $\rho = N/L^d = 1$. Both the spins and the coordinates will be considered as dynamical variables in this model. Interactions between the spins are determined by

$$I[x, S] = -\sum_{i<j} J_{ij}(x_i, x_j) W_{ij} S_i S_j - h \sum_i W_i S_i \quad , \tag{2.1}$$

with ferromagnetic coupling

$$J_{ij}(x_i, x_j) = \begin{cases} 0 & \text{if } |x_i - x_j| > R \\ J & \text{if } r < |x_i - x_j| < R \\ \infty & \text{if } |x_i - x_j| < r \end{cases} \quad , \tag{2.2}$$

giving therefore a hard core repulsion radius equal to $r/2$. As will be discussed further below, the hard core repulsive interaction is necessary for obtaining a non-trivial phase diagram, and mimics the interaction found in the dynamical triangulation model, where the minimum distance between any two spins is restricted to be one lattice spacing. For $r \to 0$, $J_{ij} = J[1 - \theta(|x_i - x_j| - R)]$. The weights $W_{ij}$ and $W_i$ appearing in Eq. (2.1) could in principle contain geometric factors associated with the random lattice subtended by the points, and involve quantities such as the areas of the triangles associated with the vertices, as well as the lengths of the edges connecting the sites. In the following we will consider only the simplest case of unit weights, $W_{ij} = W_i = 1$. On the basis of universality of critical behavior one would expect that the results should not be too sensitive to such a specific choice, which only alters the short distance details of the model.

The full partition function for coordinates and spins is then written as

$$Z = \prod_{i=1}^{N} \sum_{S_i = \pm 1} (\prod_{a=1}^{d} \int_0^L dx_i^a) \exp(-I[x, S]) \quad . \tag{2.3}$$

In the following we will only consider the two-dimensional case, $d = 2$, for which specific predictions are available from the matrix model solution. It should be clear that if the interaction range $R$ is of order one, then, for sufficiently large hard core repulsion, $r \to \sqrt{5}/2 < R$, the spins will tend to lock in into an almost regular triangular lattice. As will be shown below, in practice this crossover happens already for quite small values of $r$. The critical behavior is then the one expected for the regular Ising model in two dimensions, namely a continuous second order phase transition with the Onsager exponents. Indeed for the Ising model on a triangular lattice it is known that $J_c = \frac{1}{2}\sqrt{3}\ln 3 = 0.9514....$ On the other hand if the hard core repulsion is very small, then for sufficiently low temperatures the spins will tend to form tight ordered clusters, in which each spin interacts with a large number of neighbors. As will be shown below, this clustering transition is rather sudden and strongly first order. Furthermore, where the two transition lines meet inside the phase diagram one would expect to find a tricritical point.

In order to investigate this issue further, we have chosen to study the above system by numerical simulation, with both the spins and the coordinates updated by a standard Monte Carlo method. The computation of thermodynamic averages is quite time consuming in this model, since any spin can in principle interact with any other spin as long as they get sufficiently close together. As a consequence, a sweep through the lattice requires a number of order $N^2$ operations, which makes it increasingly difficult to study larger and larger lattices. On the other hand, we should add that we have not found any anomalous behavior as far as the autocorrelation times are concerned, which remain quite comparable to the pure Ising case.

In the course of the simulation the spontaneous magnetization per spin

$$M = \frac{1}{N}\frac{\partial}{\partial h}\ln Z|_{h=0} = \frac{1}{N} < |\sum_i S_i| > \quad , \qquad (2.4)$$

was measured (here the averages involve both the $x$ and $S$ variables, $<> \equiv <>_{x,S}$), as well as the zero field susceptibility

$$\chi = \frac{1}{N}\frac{\partial^2}{\partial h^2}\ln Z|_{h=0} = \frac{1}{N} < \sum_{ij} S_i S_j > - \frac{1}{N} < |\sum_i S_i| >^2 \quad . \qquad (2.5)$$

It is customary to use the absolute value on the r.h.s., since on a finite lattice the spontaneous magnetization, defined without the absolute value, vanishes identically even at low temperatures. In addition, in order to determine the latent heat and the specific heat exponent, we have computed the average Ising energy per spin defined here as

$$E = -\frac{1}{N}\frac{\partial}{\partial J}\ln Z|_{h=0} = -\frac{1}{JN} < \sum_{i<j} J_{ij}(x_i,x_j)\, W_{ij}\, S_i S_j > \quad , \qquad (2.6)$$

and its fluctuation,

$$C = \frac{1}{N}\frac{\partial^2}{\partial J^2}\ln Z|_{h=0} \quad . \qquad (2.7)$$

# 3  Results and Analysis

In the simulations we have investigated lattice sizes varying from $5^2 = 25$ sites to $20^2 = 400$ sites. The length of our runs varies in the critical region ($J \sim J_c$) between 1M sweeps on the smaller lattices and 100k sweeps on the largest lattices. A standard binning procedure then leads to the errors reported in the figures.

As it stands, the model contains three coupling parameters, the ferromagnetic coupling $J$, the interaction range $R$ and the hard core repulsion parameter $r$. We have fixed $R = 1$; comparable choices should not change the universality class. As we alluded previously, for small $r$ we find that the system undergoes a sharp first order transition, between the disordered phase and a phase in which all spins form a few very tight magnetized clusters. On the other hand, for sufficiently large $r$, the transition is Ising-like, between ordered and disordered, almost regular, Ising lattices (for our choice of range $R$, the transition appears to be very close to regular Ising-like for $r \approx 0.6$ and larger, see below).

A determination of the discontinuity in the average energy of Eq. (2.6) at the critical coupling $J_c$ shows that it gradually decreases as $r$ is increased from zero. Fig. 1 shows a plot of the latent heat versus $r$ at the transition point $J_c$. In general we do not expect the latent heat to vanish linearly at the endpoint, but our results seem to indicate a behavior quite close to linear. From the data we estimate that the latent heat vanishes at $r = 0.344(7)$, thus signaling the presence of a tricritical point at the end of the first order transition line. Beyond this point, the transition stays second order, as will be discussed further below. The phase transition line extends almost vertically through the phase diagram; for $r = 0$ we found on the largest lattices $J_c = 0.19(2)$, while for $r = 0.6$ we found $J_c = 0.93(3)$.

To determine the critical exponents, we resort to a finite size scaling analysis. In the following we will be mostly concerned with the values for the critical exponents in the vicinity of the tricritical point. In the case of the spin susceptibility, from finite-size scaling, we expect a scaling form of the type

$$\chi(N, J) = N^{\gamma/2\nu} \, \bar{\chi}(N^{1/2\nu} |J - J_c|) \quad . \tag{3.1}$$

To recover the correct infinite volume result one needs $\bar{\chi}(x) \sim x^{-\gamma}$ for large arguments. Thus, in particular the peak in $\chi$ should scale like $N^{\gamma/2\nu}$ for sufficiently large $N$. In Fig. 2 we show the evolution of the computed peaks in $\chi$ as a function of $\ln N$, for $r = 0.35$.

Despite the fact that the lattices are quite small, as can be seen from the graph a linear fit to the data at the tricritical point is rather good, with relatively small deviations from linearity, $\chi^2/d.o.f. = 8.8 \times 10^{-3}$. Using least-squares one estimates $\gamma/\nu = 1.27(7)$, which is much smaller than the exact regular Ising result $\gamma/\nu = 1.75$. From scaling one then obtains the anomalous dimension exponent $\eta = 2 - \gamma/\nu = 0.73(7)$. To further gauge our errors, we have computed the same exponent in the regular Ising limit, for $r = 0.6$. In this case we indeed recover the Onsager value: we find on the same size lattices and using the same analysis method $\gamma/\nu = 1.70(8)$. We also note that the shift in the critical point on a finite lattice is determined by the correlation length exponent $\nu$, namely $J_c(N) - J_c(\infty) \sim N^{-1/2\nu}$. This relationship can be used to estimate $\nu$, but it is not very accurate. From a fit to the known values of $J_c(N)$ we obtain the estimate $\nu = 1.3(2)$.

A similar finite size scaling analysis can be performed for the magnetization. Close to and above $J_c$ we expect $M \sim (J - J_c)^\beta$, and, at the critical point on a finite lattice, as determined from the peak in the susceptibility (which incidentally is very close to the inflection point in the magnetization versus $J$), $M$ should scale to zero as $M_N(J_c) \sim N^{\beta/2\nu}$. In Fig. 3 we show the magnetization $M$ computed in this way for different size lattices close to the tricritical point. In spite of the larger errors the results again clearly exclude the pure Ising exponents, and give $\beta/\nu = 0.30(10)$, to be compared to the exact regular Ising result $\beta/\nu = 0.125$. A similar analysis in the pure Ising limit (more precisely, for $r = 0.6$) gives $\beta/\nu = 0.15(7)$.

The results for the peak in the Ising specific heat $C$ at the tricritical point as a function of lattice size $L$ are shown in Fig. 4. One expects the peak to grow as $C \sim N^{\alpha/2\nu}$, but the absence of any growth implies that $\alpha/\nu < 0$ (a weak cusp in the specific heat). If we insist on fitting the peak in the specific heat to a power of $N$, we get $\alpha/\nu \approx -0.11(5)$, a negative value due to the decrease of the peak with increasing system size. On the other hand we should add that, in general close to a critical point, the free energy can be decomposed in a regular and a singular part. In our case the singular part does not seem to be singular enough to emerge above the regular background, leading to an intrinsic uncertainty in the determination of an $\alpha < 0$, and which can only be overcome by determining still higher derivatives of the free energy with respect to the coupling $J$. A better approach would seem therefore to determine the correlation length exponent $\nu$ instead, and use scaling to relate it to $\alpha = 2 - 2\nu$. In the regular Ising case one has in a finite volume a logarithmic divergence $C \sim \ln N$ (and $\alpha/2\nu = 0$), and we indeed see such a divergence clearly for $r = 0.6$, which corresponds to the almost regular triangular Ising case.

One can improve on the estimate for $J_c$ by considering the fourth-order cumulant

|              | $\gamma/\nu$ | $\beta/\nu$ | $\alpha/\nu$ | $\nu$   |
| ------------ | ------------ | ----------- | ------------ | ------- |
| This work    | 1.27(7)      | 0.30(10)    | < -0.11(5)   | 1.3(2)  |
| Matrix model | 1.333...     | 0.333...    | -0.666...    | 1.5     |
| Onsager      | 1.75         | 0.125       | 0.0          | 1.0     |
| Tricritical Ising | 1.85    | 0.075       | 1.60         | 0.555...|

Table 1: Estimates of the critical exponents for the random two-dimensional Ising model, as obtained from finite size scaling at the tricritical point.

[9]

$$U_N(J) = 1 - \frac{<m^4>}{3<m^2>^2} \quad , \tag{3.2}$$

where $m = \sum_i S_i/N$. It has the scaling form expected of a dimensionless quantity

$$U_N(J) = \bar{U}(N^{1/2\nu}|J - J_c|). \tag{3.3}$$

The curves $U_N(J)$, for different and sufficiently large values of $N$, should then intersect at a common point $J_c$, where the theory exhibits scale invariance, and $U$ takes on the fixed point value $U^*$. We have found that indeed the curves meet very close to a common point, and from the intersection of the curves for $N = 25$ to 400 we estimate $J_c = 0.48(1)$, which is consistent with the estimate of the critical point derived from the location of the peak in the magnetic susceptibility. We also determine $U^* = 0.47(4)$, to be compared to the pure Ising model estimate for the invariant charge $U^* \approx 0.613$ [10].

One can estimate the correlation length exponent $\nu$ from the scaling of the slope of the cumulant at $J_c$. For two lattice sizes $N, N'$ one computes the estimator

$$\nu_{eff}(N, N') = \frac{\ln[N'/N]}{2\ln[U'_{N'}(J_c)/U'_N(J_c)]} \quad , \tag{3.4}$$

with $U'_N \equiv \partial U_N/\partial J$. Using this method, we find $\nu = 1.3(3)$.

In Table I we summarize our results, together with the exponents obtained for the two-matrix model [1], for the Onsager solution of the square lattice Ising model, and for the tricritical Ising model in two dimensions [8]. As can be seen, the exponents are quite close to the matrix model values (the pure Ising exponents seem to be excluded by several standard deviations).

## 4 Conclusions

In the previous sections we have presented some first results for a random Ising model in flat two-dimensional space. The model reproduces some of the features of a model for dynamically triangulated Ising spins, and in particular its random nature, but does not incorporate any effects due to curvature. Due to the non-local nature of the interactions of the spins, only relatively small systems could be considered so far, which is reflected in the still rather large uncertainties associated with the exponents. Still a rich phase diagram has emerged, with a tricritical point separating first from second order transition lines. We have localized the tricritical point at $J_c = 0.48(1)$ and $r = 0.344(7)$. The thermal and magnetic exponents determined in the vicinity of the tricritical point (presented in Table I) have been found to be consistent, within errors, with the matrix model solution of the random Ising model. Our results would therefore suggest that matrix model solutions can also be used to describe a class of annealed random systems in flat space.

**Acknowledgements**

The numerical computations were performed on facilities provided by the San Diego Supercomputer Center (SDSC), the University of California at Irvine, and by the Texas National Research Laboratory Commission through grants RGFY9166 and RGFY9266.

Figure 1: Latent heat along the first order transition line, plotted against the hard core repulsion parameter $r$. The tricritical point is located where the latent heat vanishes.

Figure 2: Peak in the magnetic susceptibility, $\chi_{max}$, versus the number of Ising spins $N$, for fixed hard core repulsion parameter $r = 0.35$.

Figure 3: Finite size scaling of the magnetization at the inflection point, $M_{inf}$, versus the total number of Ising spins $N$, for fixed hard core repulsion parameter $r = 0.35$.

Figure 4: Plot of specific heat $C$ versus ferromagnetic coupling $J$ at $r=0.35$, showing the absence of a growth in the peak with increasing lattice sizes, in contrast to the behavior of the magnetic susceptibility. The errors (not shown) are smaller than the size of the symbols.

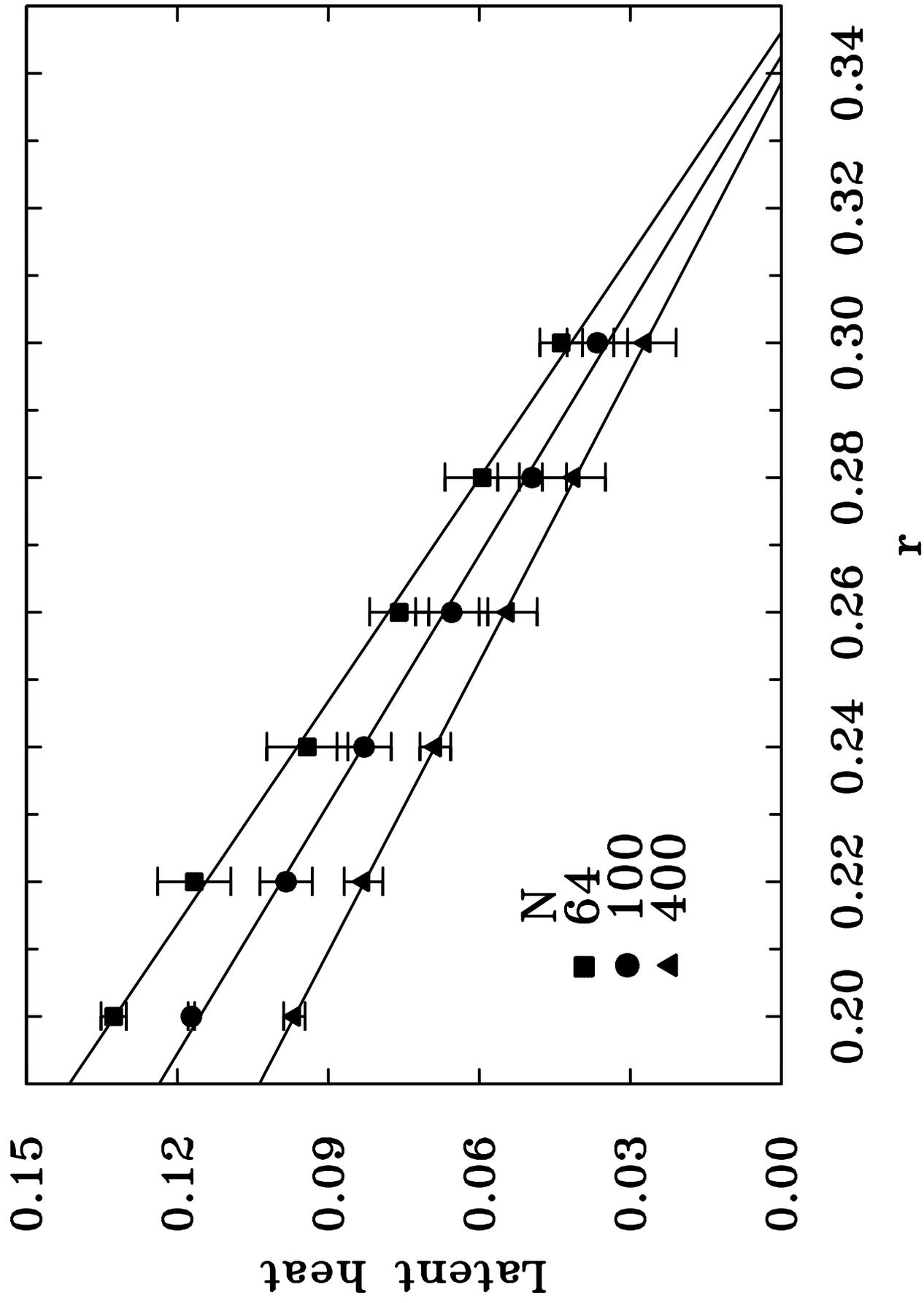

Fig.1

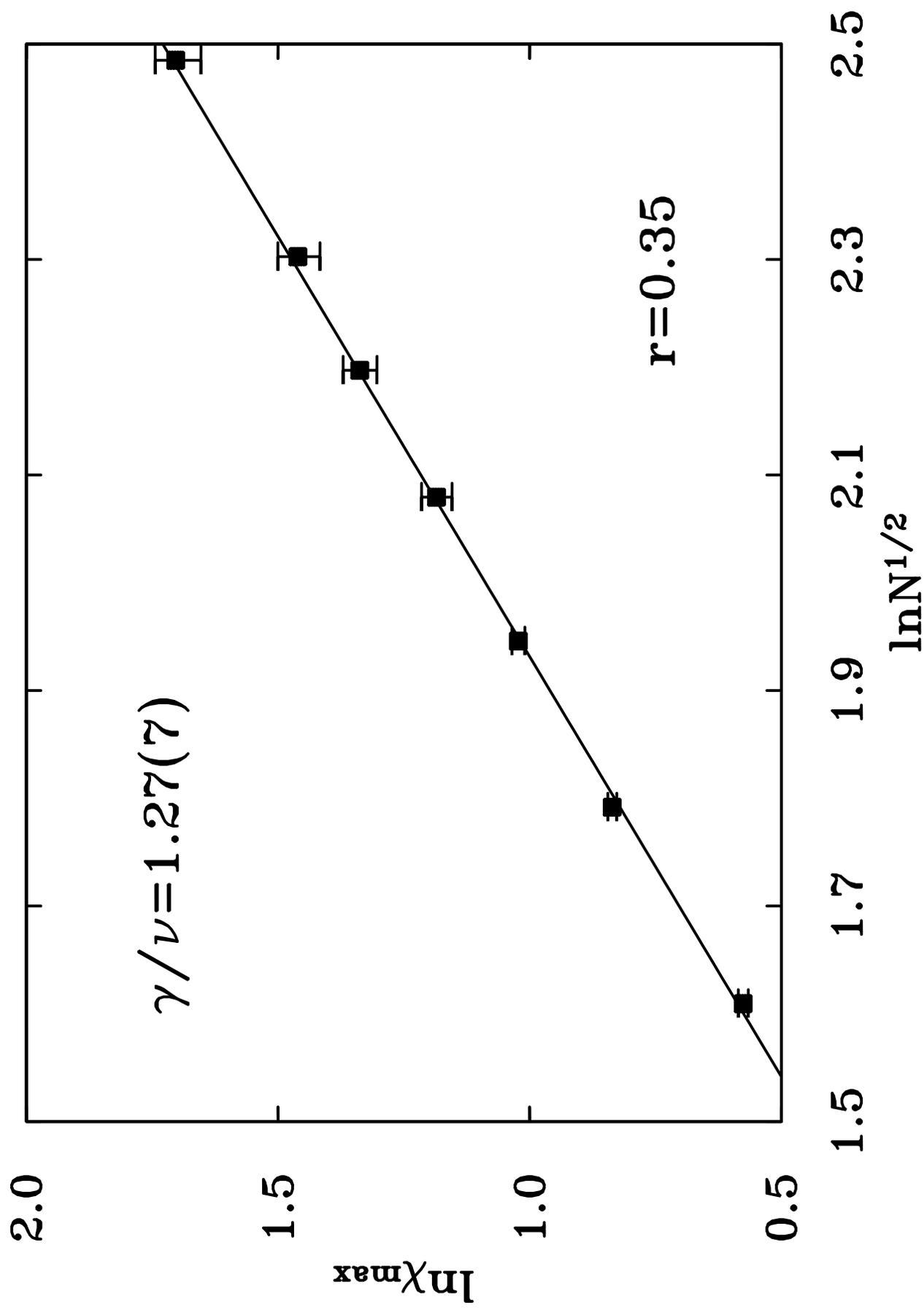

Fig.2

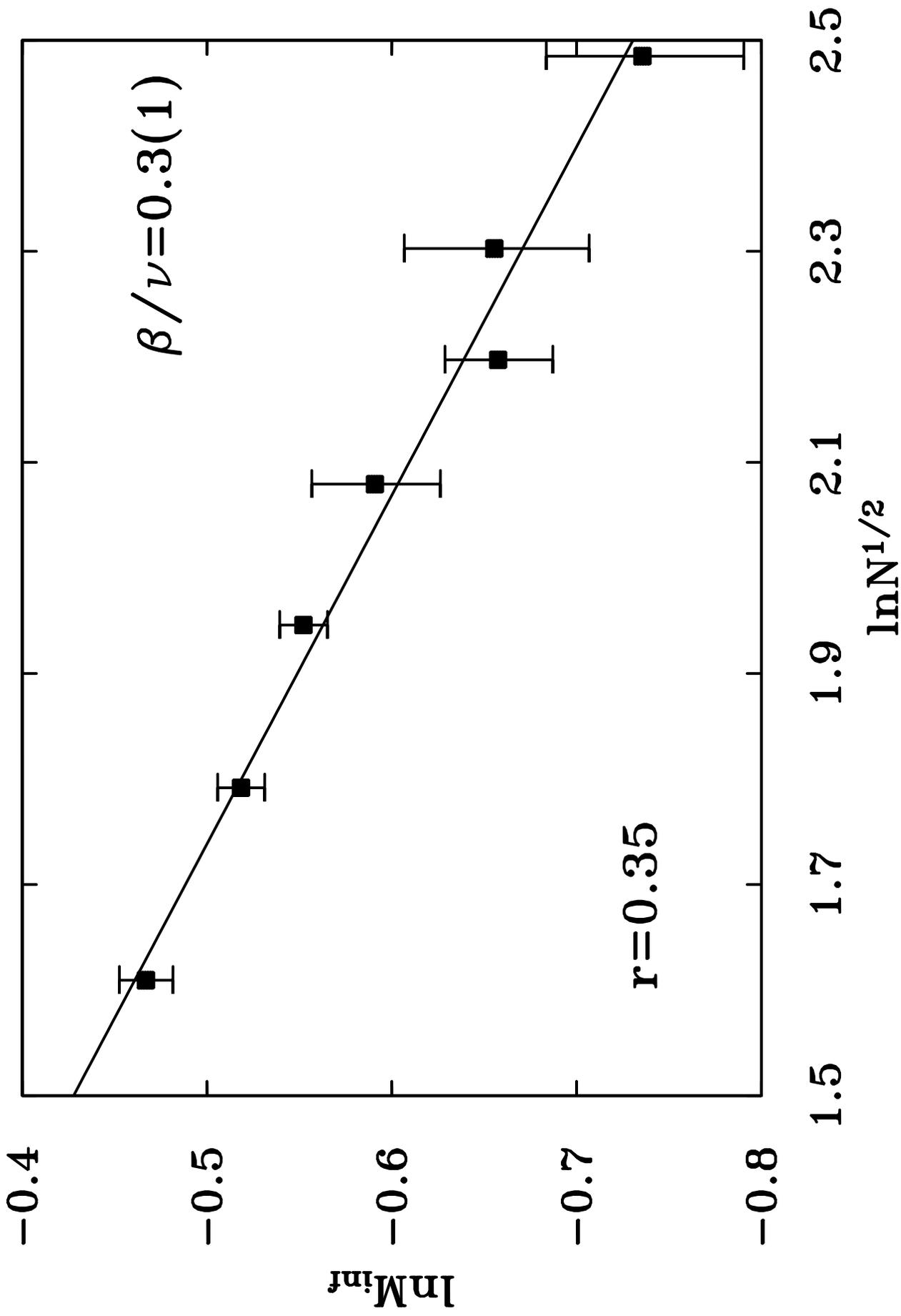

Fig.3

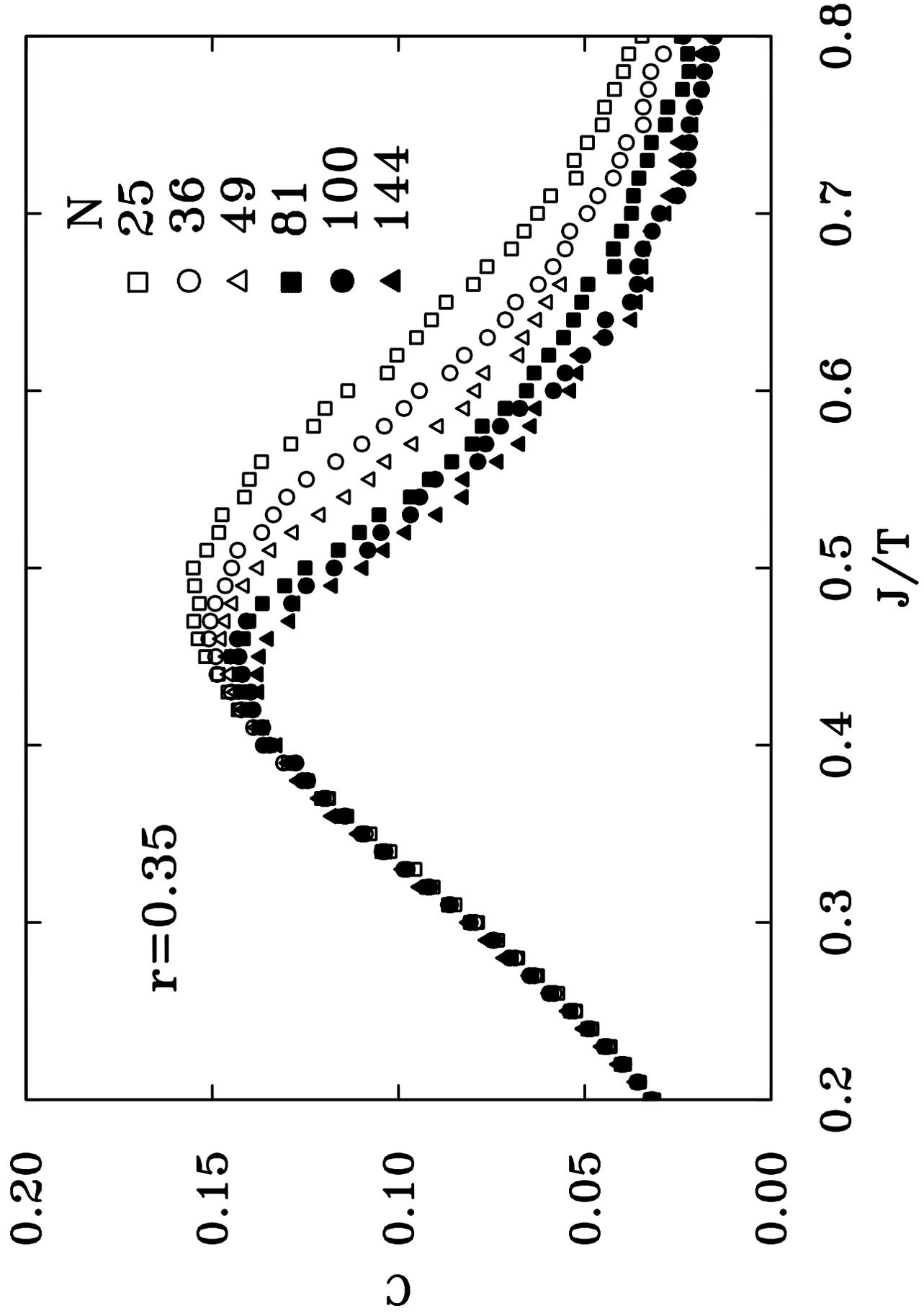

Fig.4